# A Real-time and Hardware Efficient Artfecat-free Spike Sorting Using Deep Spike Detection

Xiaoyu Jiang, Student, IEEE, Tao Fang, Student, IEEE, and Majid Zamani, Member, IEEE

*Abstract*: Spike sorting is a valuable tool in understanding brain regions. It assigns detected spike waveforms to their origins, helping to research the mechanism of the human brain and the development of implantable brain-machine interfaces (iBMIs). The presence of noise and artefacts will adversely affect the efficacy of spike sorting. This paper proposes a framework for low-cost and real-time implementation of deep spike detection, which consists of two one-dimensional (1-D) convolutional neural network (CNN) model for channel selection and artefact removal. The framework utilizes simulation and hardware layers, and it applies several low-power techniques to optimise the implementation cost of a 1-D CNN model.  A compact CNN model with 210 bytes memory size is achieved using structured pruning, network projection and quantization in the simulation layer. The hardware layer also accommodates various techniques including a customized multiply-accumulate (MAC) engine, novel fused layers in the convolution pipeline and proposing flexible resource allocation for a power-efficient and low-delay design. The optimized 1-D CNN significantly decreases both computational complexity and model size, with only a minimal reduction in accuracy. Classification of 1-D CNN on the Cyclone V 5CSEMA5F31C6 FPGA evaluation platform is accomplished in just 16.8 microseconds at a frequency of 2.5 MHz. The FPGA prototype achieves an accuracy rate of 97.14% on a standard dataset and operates with a power consumption of 2.67mW from a supply voltage of 1.1 volts.  An accuracy of 95.05% is achieved with a power of 5.6mW when deep spike detection is implemented using two optimized 1-D CNNs on an FPGA board.

## I. INTRODUCTION

Extracellular recordings have been widely used to monitor neuronal activity by implanting multi-electrodes in the cortex and capturing multidimensional neural data. A processing step, known as spike sorting shown in Fig.1(a), is necessary to separate the multi-unit neural activities and assign the captured spikes to their originating neurons [1-9]. Spike sorting is an invaluable research tool applied in implantable brain-machine interface (iBMI) research for studying and decoding neural signals from different brain regions and understanding the mechanisms of the brain. It is extremely beneficial in design and development of various applications such as identifying the optimal patterns and parameters to condition diseases by artificially modulating irregular or faulty electrical impulses [10], realizing a communication bridge for control of assistive devices for patients with damaged sensory/motor functions such as hand prosthesis [11], and stimulating a particular pathway for biological functionality regularization [12].

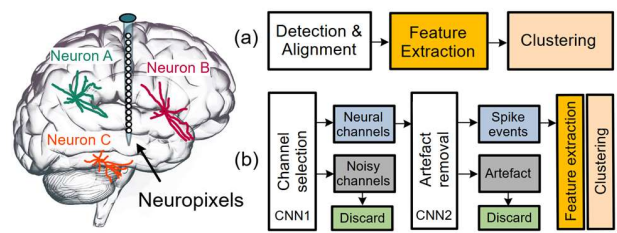

Fig. 1. Recording activities from neurons A, B and C using Neuropixels electrode arrays. (a) Spike sorting process for determining single unit activity using conventional methods. (b) Spike events monitoring using deep spike detection which incorporates two 1-D convolutional neural networks (CNNs), CNN1 and CNN2 for channel selection and artefact removal respectively. This is known as artefact-free spike sorting.

The recent trend in brain sensing is about the utilization of high-channel count neural interfaces that include tens of thousands of sensing probes [13]-[14]. In such recording set-ups, the extracellular recordings are typically severely contaminated by artefacts and various noise sources, rendering the separation of multi-unit neural recordings an immensely challenging task. Therefore, removing artefacts and noise from neural events is not only crucial but also directly enhances spike sorting performance and classification accuracy.

This motivates a new paradigm in spike events detection for artefact-free sorting in high-channel count recording, called deep spike detection shown in Fig.1(b). Deep spike detection incorporates two convolutional neural networks (CNNs) into the conventional spike processing framework to identify the active neural channels and eliminate artefacts from those selected channels. It identifies and extracts distinctive spike and artefact characteristics from the input channels, enabling the selective elimination of artefacts from extracellular recordings. In 2019, Saif-ur-Rehman et al proposed SpikeDeeptector [15], to detect and track channels containing neural data. SpikeDeeptector employs a semi-automatic approach to generate pseudo-labels, which involves visual inspections [15]. The training dataset was derived from a single subject

X. Jiang, and M. Zamani are with the School of Electronics and Computer Science, University of Southampton, Southampton, SO17 1BJ UK, (e-mail: xj1u23@soton.ac.uk; m.zamani@soton.ac.uk; km3@ecs.soton.ac.uk).

T. F. was with  Engineering and Applied Technology Research Institute, Fudan University, 220 Handan Road, Shanghai, 200433, China (e-mail: 19210860039@fudan.edu.cn).

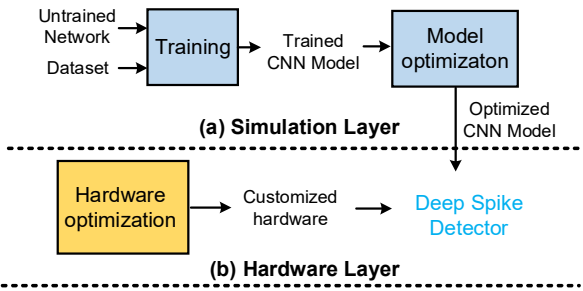

Fig.2. Two-layer optimization framework. (a) Simulation layer. (b) Hardware layer. The untrained model is fed to the input of the simulation layer for initial optimizations. The optimized model is sent to the hardware layer for customized hardware design.

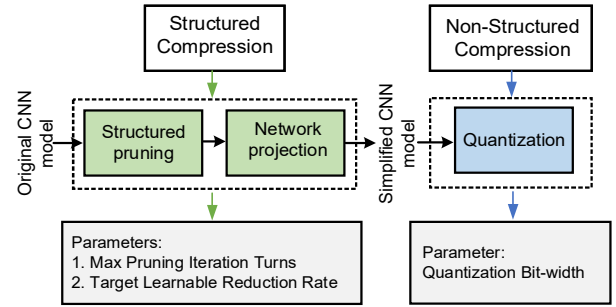

Fig.3. Simulation layer optimization procedures. Structured compression techniques including pruning and network projection simplify the original model. Non-structured compression further optimizes the 1-D CNN by minimizing the bit-width of the parameters.

comprising 1.56 million feature vectors that represent batches of waveforms from one channel, each with dimensions of 48()×20 samples. In the first CNN, if at least one of the waveforms represented a spike, the whole batch was classified as neural activity. Authors reported an overall accuracy of 97.2% is achieved with SpikeDeeptector [15]. Notably, the model could generalize across different brain regions, subjects, species and electrode types. The second CNN is called 'background activity rejector' (BAR) that takes signals selected by SpikeDeeptector and discards background activity by classifying a given signal as spike or noise. BAR achieves an accuracy of 92.3%. In 2023, Okreghe et al [16] also proposed a solution for real-time improved channel selection and artefact removal. The method in [16], also works with two CNNs, one for the selection of channels and one for artefact removal. Similar to [15], the proposed work in [16] trained with 1.61million labelled feature vectors to create a robust network. The first CNN labels an input batch (48×20) as 'neural' if it monitors at least one spike waveform. Like BAR [15], the second CNN is provided with a single feature vector (48×1) and discards non-neural events, and it achieves an overall accuracy of 92.3%. The issue with the existing deep spike detectors [15]-[16] is that they are implemented in software and can not be utilized for real-world implantable brain processing. There is therefore a need for hardware implementation of a deep spike detector that uses extremely low power consumption to meet the power-area requirements of implantable devices. Following our previous work in [16], this paper proposes a low-cost framework to implement deep spike detection shown in Fig.2. The first layer of the proposed framework focuses on the optimization techniques to reduce the number of parameters and to diminish the memory space. A compact model with 210 bytes of memory size is achieved using structured pruning, network projection and quantization in the simulation layer. The efforts in the second layer of the proposed framework are mainly made to design customized and real-time hardware for the optimized 1-D CNN from the simulation layer. The proposed optimization framework significantly reduces the overall power consumption and classification latency of 1-D CNN in hardware. An accuracy of 95.05% is achieved with a power of 5.6mW when the deep spike detector is implemented using two optimized 1-D CNNs on Cyclone V 5CSEMA5F31C6 FPGA. The rest of the paper is structured as follows: Section II describes the simulation layer of the proposed optimization framework. The customized hardware techniques are presented in Section III, followed by an FPGA and implementation results in Section IV. Finally, Section V makes some concluding remarks.

## II. PROPOSED OPTIMIZATION FRAMEWORK

In this section, adopted techniques for model optimization are explained, together with the optimization procedures. Also, the optimization results are demonstrated.

### A. Simulation Layer

Fig.3 shows the procedures of model optimization in the simulation layer. Structured and non-structured compression techniques including pruning, network projection and quantization are applied on the original 1-D CNN model.

Structured pruning refers to removing entire filters from the network, thus changing the structure of the model. For example, in the convolutional layers of a CNN model, reducing the filter numbers from 4 to 2 will result in a dimensionality reduction for subsequent feature maps from 4 channels to 2 channels. This approach can significantly affect the model size and achieve acceleration with the help of standard hardware [17]-[18]. Network projection [19] refers to mapping the network from a higher dimension into a lower dimension. In this process, the learnable parameters are projected into the subspace while maintaining the highest variance in neuron activations based on the principal component analysis (PCA) on the neuron activations. Network projection replaces a layer with a subnetwork of smaller layers with fewer parameters. Suitable network projection can reduce the number of learnable parameters and retain a high prediction accuracy at the same time.

Quantization [20] also converts the parameters of CNN, from high precision pseudo-continuous values to discrete values of low precision. This technique reduces the bit-width of each parameter and the overall size of the CNN model. A common approach of quantization is to convert the parameters from a floating point 32 bits to INT8 format or even fewer bits. Due to the noise tolerance of CNN, the network after quantization is able to maintain its performance compared to the unquantized version.

The computation of floating point is time and resource consuming because it needs the calculation of three parts under this format: sign, exponent and mantissa. In comparison, a fixed-point value such as INT8 computations can be easily performed by full adders which is more efficient for embedded or customized hardware.

In each optimization technique, there is a parameter that needs to be modified, which results in different optimization outcomes. In the simulation layer, the parameters are max pruning iteration turns, target learnable reduction rate and quantization bit-width which corresponds to the structured pruning, network projection and quantization as shown in Fig. 3. Aggressive parameter optimization can result in a more compressed model, however this will result in a greater possibility of accuracy loss. For example, the max pruning iteration controls the number of removed convolutional filters. If too many filters are removed, the accuracy of the CNN model will drop dramatically. To avoid this situation, the accuracy of CNN models is examined after each optimization run.

*B. Optimization Procedures in Simulation Layer*

The first step of model optimization is to construct a CNN, this is regarded as the original or unoptimized model. The test dataset (see Section IV. B) is split into three parts for training, validation and test datasets with the ratio of 0.7, 0.15, 0.15. The proposed deep spike detection algorithm in this paper works with two 1-D CNNs, one for the selection of channels and one for artefact removal similar to [13]-[14]. Therefore, a similar CNN (66×1) is plugged into the two stages to perform deep spike streaming over time, where 66 is the sample waveform segment for a time duration of 2.5 ms. In the channel selection stage, a down-sampling factor of 10 is considered to reduce the number of samples for in-channel activity analysis. This helps to monitor a larger segment of data with a much smaller number of samples and is also aligned with the sparse behaviour of neural activities in the recorded data. Considering the down-sampling factor, the first CNN accepts a batch of (66×1) while it monitors spiking activities in a data segment with 660 samples.

A 1-D CNN model is therefore constructed for the optimization procedures. The input layer is a 1-D Array with a length of 66 that shows the number of samples per data segment, and can be used in channel selection and artefact removal stages shown in Fig. 1(b).

50 convolution kernels are utilized in each convolution layer. In the first convolutional layer (referred to as Conv1 Layer), the kernel size is 1×3 for a single input channel. In the next two convolutional layers (Conv2 and Conv3 Layers), the spatial size of the kernel remains 1×3, but the number of channels increases to 50. ReLU layers are placed after each convolutional layer and fully connected layer as activation functions to improve the learning ability of the model by increasing its non-linearity. Maxpooling layers are also utilized to reduce the amount of calculation, but maintain the important features and information. Fully connected (FC) layer is the final layer in the 1-D CNN consisting of 750 neurons.

Mini-batch gradient descent with momentum (Mini-batch SGDM) is used to solve the optimization problem. With a learning rate that starts at 0.01 and is adjusted piecewise, increasing by 10 for every 5 training epochs. The maximum number of epochs and the batch size are set to 200 and 256 respectively. Using L2 regularization, grid search was conducted from 0 to 5 with a step size of 0.2, the optimal value was 1.8. By tracking the validation error, an early stopping criterion was used to avoid overfitting on the test data at each epoch. The training is stopped after six consecutive epochs if the error rises or stays the same. Lastly, drop-out regularization was applied. After training, a CNN model with 99.43% classification accuracy, 17553 learnable parameters and 68Kbytes of memory is generated. This CNN model is used as the original or unoptimized model for the optimization procedures.

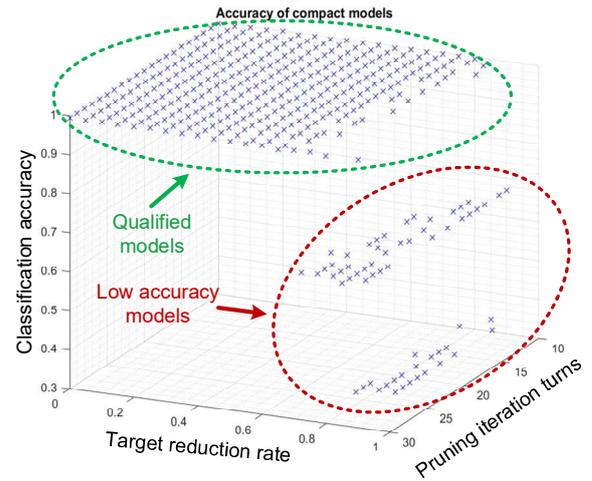

Fig.4. Classification accuracy of the compressed models in Step 1.

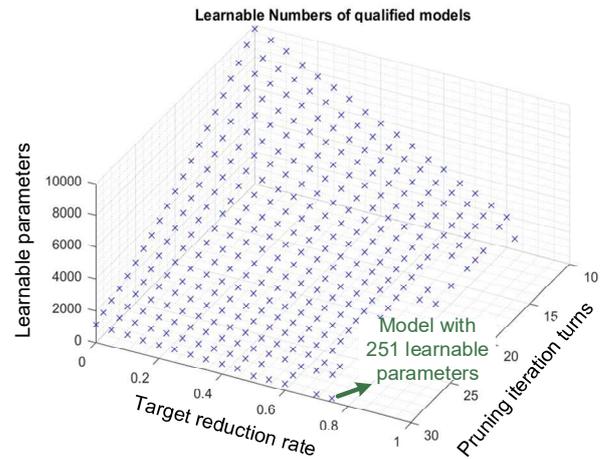

Fig.5. Number of learnable parameters in compressed models from Step 1. The green arrow shows a model with 251 parameters.

The optimization procedures will begin with the structured compression including pruning and network projection as shown in Fig. 3. A validated and examined model after structured pruning is then sent to the non-structured compression which applies the quantization for further implementation cost reduction. Two evaluation metrics are used to validate the optimization steps. The first evaluation metric is the classification accuracy on the testing dataset (see Section IV.B). The second metric is the number of learnable parameters of the optimized model. A model with the least learnable number and maintaining an accuracy very close to 100% is the most ideal one. The following steps are carried out to find the final optimized model.

Step 1) Various compressed models are investigated in the structured compression stage, by running pruning and network projection. The models with classification

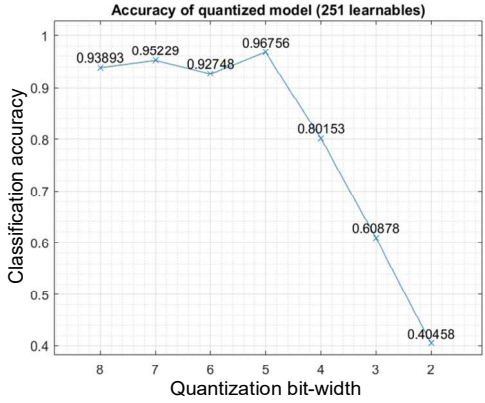

Fig.6. Classification accuracy of the model with 251 learnable parameters versus quantization bit-width.

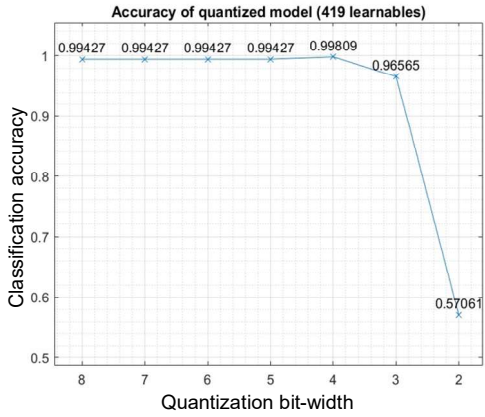

Fig.7. Classification accuracy of the model with 419 learnable parameters versus quantization bit-width.

accuracy above 99% with the least number of learnable parameters are retained.

Step 2) Perform different quantization bit-width on the model found in Step1. If the model fails to maintain stability following quantization, consider utilizing a version of the model with a marginally increased number of parameters in Step 1, continuing this process until stability is achieved with at least one quantization bit-width.

Step 3) Select the model with the least quantization bit-width while maintaining high accuracy.

It is important to note that the maximum number of pruning iterations is set to 30, the target reduction of the learnable rate is specified within the range of 0 to 0.9, and the quantization bit-width is taken into account within the limits of 8 to 2.

Fig.4 and Fig.5 show the outcome of optimization procedures outlined in Step 1. After discarding all the unqualified models shown with a red circle in Fig.4, Fig.5 shows the number of learnable parameters in the qualified models. A model with the 251 learnable parameters is sent to the quantization step. The selected model with 251 learnables from Step1 shows poor stability during the quantization step, shown in Fig.6. The accuracy of this model will drop to 93.90% when quantized to 8-bits. The accuracy of the optimized CNN demonstrates instability and will drop more if quantized to fewer bits, therefore this model is quite vulnerable to quantization. Another model with 419 learnable parameters and 99.62% accuracy is selected for Step2. The same quantization process is performed on this model. As shown in Fig.7, this model maintains a high classification accuracy of 99.43% when quantized to 8-bits. This model still maintains its performance even when quantized to 4-bits. Its accuracy will drop to 96% when quantized by 3-bits.

TABLE I
Comparison between the original and the optimized 1-D CNNs.

| Layers | Original CNN | Optimized CNN | Sub-Layers |
|---|---|---|---|
| **Conv1 Layer** | W: 1×3×1×50<br>B : 1×1×50<br>A: 1×66×50×1 | W: 1×3×1×1<br>B : 1×1×1<br>A : 1×66×1×1 | **Conv1**: Conv Layer |
| | | W: 1×1×1×10<br>B : 1×1×10<br>A : 1×66×1×1 | **Conv1**: Projection Out Layer |
| **Conv2 Layer** | W: 1×3×50×50<br>B : 1×1×50<br>A: 1×64×50×1 | W: 1×1×10×1<br>B : 1×1×1<br>A : 1×66×1×1 | **Conv2**: Projection In Layer |
| | | W: 1×3×1×1<br>B : 1×1×1<br>A : 1×64×1×1 | **Conv2**: Conv Layer |
| | | W: 1×1×1×10<br>B : 1×1×10<br>A : 1×64×10×1 | **Conv2**: Projection Out Layer |
| **Conv3 Layer** | W: 1×3×50×50<br>B : 1×1×50<br>A: 1×30×50×1 | W: 1×1×10×1<br>B : 1×1×1<br>A : 1×32×1×1 | **Conv3**: Projection In Layer |
| | | W: 1×3×1×2<br>B : 1×1×2<br>A : 1×30×2×1 | **Conv3**: Conv Layer |
| | | W: 1×1×2×10<br>B : 1×1×10<br>A : 1×30×10×1 | **Conv3**: Projection Out Layer |
| **FC Layer** | W: 3×750<br>B : 3×1<br>A: 3×1 | W: 2×150<br>B : 2×1×<br>A : 2×1 | FC: Projection In Layer |
| | | W: 3×2<br>B : 3×1×<br>A : 3×1 | FC: Projection Out Layer |

In comparison, the former model has the least learnable parameters but is unable to resist any quantization, the latter model has slightly more parameters but maintains its performance after 4 and 3-bits quantization. The latter is a more suitable model for deployment for the hardware layer. As a result, a 4-bit quantization for the model with 419 learnables is an ideal option, where the quantized model can shrink the memory requirement to a great extent and maintain good classification performance at the same time.

The compact CNN model after structured and non-structured compressions is generated by max pruning iterations of 30, target learnable reduction rate of 0.6 and the quantization bit-width set to 4-bits. This CNN has 419 total learnable parameters, with a classification accuracy of 99.81%. Table I demonstrates the main layers of the non-optimized and the optimized CNNs. The weights (W), bias (B) and activations (A) are shown in the form of SSCB (spatial, spatial, channel, batch).

From Table I, it can be seen that the channel number decreased from 50 to 10. This is due to the fact that the structured pruning removed plenty of filters from the original model. Also, due to the use of network projection, the convolution layers and fully connected layers are all replaced by smaller sublayers. For example, the original Conv2 layer in the non-optimized model is replaced by three sublayers.

Utilization of structured pruning and network projection resulted in a great compression in a 1-D CNN model size, the learnable number decreased from 17553 to 419 (i.e., 41X compression rate). Also, considering the 4-bits quantization performed on the model, each parameter only occupies 4 bits instead of 32 bits. This resulted in a memory decrease from 70212 bytes to 210 bytes (i.e., 334X memory compression). However, the use of network projection results in an increase in layer numbers from 12 to 18, where each layer is expanded into several smaller sub-layers detailed in Table I.

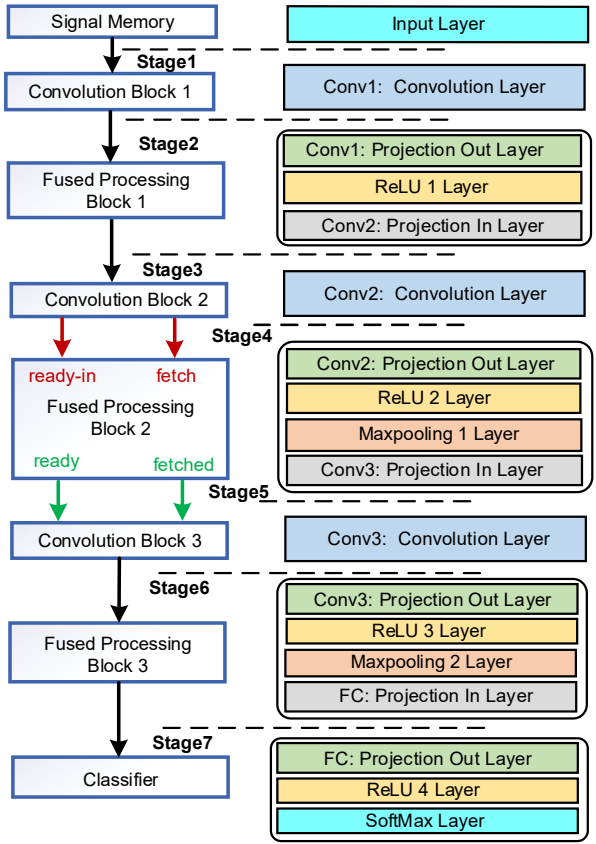

Fig.8. Architecture of the proposed 1-D CNN using hand-shaking protocol.

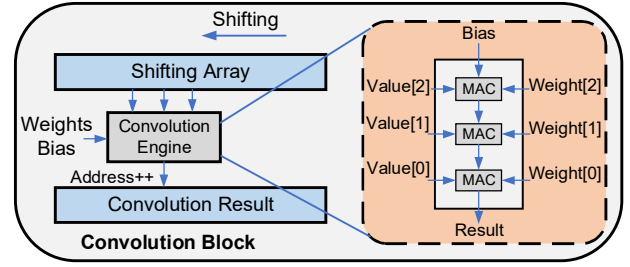

Fig.9. Architecture of convolution block. Three MAC units are used in the convolution engine to perform shift-and-convolution per array.

## III. HARDWARE LAYER OPTIMIZATION

After applying the optimization techniques in the simulation layer, a customized hardware is designed to execute the 1-D CNN model. This section will demonstrate the main techniques used in the hardware design.

### A. The 1-D CNN Model Architecture

The hardware design adopted the idea of pipelining consists of several hardware blocks connected in series to perform classification, and the model diagram of the proposed network model is shown in Fig. 8. The number of input channels in this model is set to 66 for channel selection and the artefact removal stage. The equivalent model of the fused processing blocks (see Section III. C) are shown in more details on the right column.

The design in Fig. 8 doesn't have a global control unit, therefore handshaking protocols and signals are used to control the pipeline running between the blocks. The data processing thereafter is driven only by local handshaking events without a global synchronous clock, eliminating any timing uncertainty such as skew and jitter associated with clock distribution. The handshaking protocol between the modules is reliably governed by four signals including ready, ready_in, fetch and fetched. The ready and fetched signals are used for communications with the upcoming block, and the ready_in and fetch signals control the current block as shown in Fig. 8. Consequently, each self-timed block functions at its own pace consistently, delivering optimal performance that adjusts dynamically to varying operating conditions. This also aids in effectively reducing the operational cost of the proposed pipeline [21].

Accessing external memory is an energy-consuming activity requiring more operations and extra interactions, conversely, accessing local memory is quicker and more efficient [22]. To avoid unnecessary data movement operations, the overall memory of the complete 1-D CNN model is separated into several parts and distributed across each hardware block. Therefore, each hardware block has its own memory to store parameters belonging to its responsible layers. This can be seen as distributed local memory for one convolution layer to store the weights and bias data for convolution operations.

### B. Convolution Engine

The designed and utilized convolution block in the 1-D CNN is shown in Fig. 9. There are three main components in this block including the convolution engine, shifting array and array to store convolution results. Each convolution layer has its own block for accelerated classification.

The convolution engine involves three multiply-accumulate (MAC) units connected in series due to the convolution kernel size as shown in Fig. 9. The convolution kernel slides through the entire array based on the stride size, each time a matrix multiplication is performed between the convolution kernel and the shifting array.

Each MAC consists of a multiplier and an adder [18]. The convolution process begins with selecting three elements from the shifting array, and convolution is performed with the kernel's coefficients. When a convolution calculation is performed, the result is written into a result array, with its address (or index) incremented by one for the next convolution operation. Also, the shifting array will shift by one time to simulate the sliding window of filters. For a shifting array with a length of n, (n-2) times of shifting are needed to complete the entire convolution process.

### C. Fused Layers and Processing

For the optimized CNN model from the simulation layer, there are plenty of matrix projection operations in the pipeline shown in Fig. 8. Inspired by [23], the projection of a larger matrix into a smaller matrix occurs prior to the convolution operation. Following the convolution, the smaller matrix is transformed back into a larger matrix and subsequently passes through the succeeding layers, including ReLU and Maxpooling, until the commencement of the next convolution. A High-bandwidth data transfer (66×10) is observed between the output of one projection out layer and the next projection in layer in Fig. 10 (a). A fused processing block is designed in the CNN pipeline to merge several layers together, changing the data flow and computation pattern to avoid the high-bandwidth data transfer as shown in Fig. 10 (b).

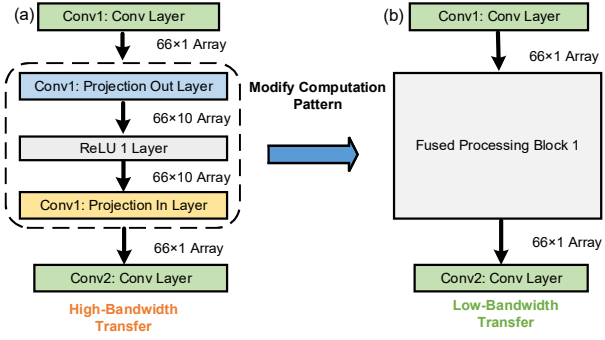

Fig.10. (a) High-bandwidth data transfer. (b) Modifying computation pattern using fused processing block. This avoids loading the entire feature map before computing, equal to low-rank matrix computations.

Taking the fused processing block in Fig. 10 as an example, this block needs to manage the tasks of projection out, ReLU and projection in layers. The process begins with the projection out layer in Fig. 11 which is expressed as $c = w^T \cdot a + b^T$, where $w$, $a$ and $b$ are the weights, input array and bias in the projection out layer. $T$ also refers to the matrix transpose. It should be noted that the matrix $c$ has m rows and n columns as $c_{ij} = w_i a_j + b_i$ $(1 \ll i \ll m, 1 \ll j \ll n)$. The matrix $d$ after ReLU is presented as $d = Max(0, c)$ with m rows and n columns, the detailed version is written as $d_{ij} = Max(0, w_i a_j + b_i)$ $(1 \ll i \ll m, 1 \ll j \ll n)$. The final layer in Fig. 10 (a) and Fig.11 is projection in layer that projects the matrix $d$ onto $e = w'^T \cdot d + b'$. Matrix $e$ has 1 row and n columns with each element defined as $e_j = b' + \sum_{i=1}^{m} w'_i \cdot Max(0, w_i a_j + b_i)$ $(1 \ll j \ll n)$, where $w'$ and $b'$ are the weights and the biases of the projection in layer.

Throughout the fusion process, each element in the output array $e$ is determined by only one element from the input array $a$. That is a one-to-one mapping from the input array to the output array based on the known weights and biases.

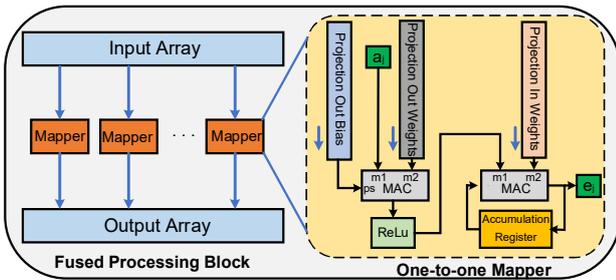

Fig.11. Architecture of the fused processing block. The fusion is performed using mappers detailed in the yellow box.

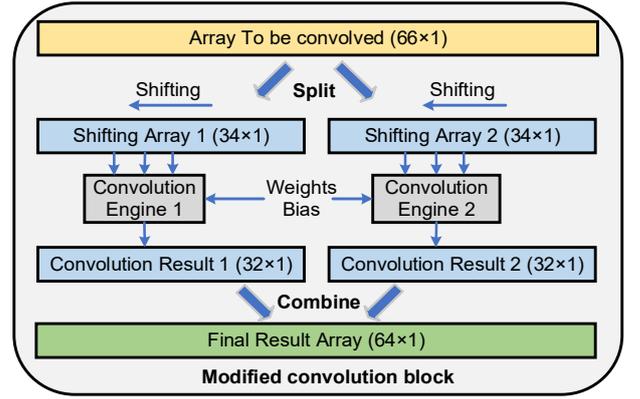

Fig.12. Performing two convolutions in parallel to adjust the delay in the convolution engine. Example of an array (66×1) splitting into two arrays (33×1) for parallel convolution is annotated in the figure.

Fig.11 demonstrates the details of a one-to-one mapper and the structure of a fused processing block. The weights ($w$) and bias ($b$) during projection are all stored in three shifting arrays and are shifted by once in a clock cycle. $m1$, $m2$ and $ps$ represent the multiplier, multiplicand and partial sum ports of a MAC unit. The mapping operation is performed in $m$ clock cycles, where $m$ represents the number of rows in the matrix which is 10 in this work. The introduction of this component helps fuse several layers together and avoids the high bandwidth data transfer between blocks.

Because the mapping operation only depends on the internal parameters and the input array, several mappers can work independently at the same time increasing the parallelism of computation. It is important to note that certain fused processing blocks incorporate Maxpooling layers, resulting in a shift from a one-to-one mapping to a two-to-one mapping. Nevertheless, the overall architecture remains unchanged, with the addition of a 2-input comparator.

### D. Roof-line Model and Resource Allocation

Considering the roof-line model [24], a resource allocation strategy is adopted to introduce a balance between workload and computational power. The golden point in the roof-line model is satisfied when a block is able to use allocated computing power and continuously work on data processing without an interval (i.e., to avoid bucket effect in the pipeline). To achieve this objective, resource allocation must vary across different blocks due to the differing quantities of tasks assigned to each.

First, a block with a relatively low workload is identified. Then the processing time of the block for an input batch is calculated as a reference. Finally, the amount of workload in the block is adjusted to meet the time specification of the design. Conv3 is set as a benchmark, requiring approximately 30 clock cycles to finalize the data processing. The resources assigned to the other blocks are adjusted based on this reference to ensure a balanced time delay. For example, 3 MAC units are used in Conv2 to perform the convolution, however it will take 64 clock cycles to complete the process which means a delay in the pipeline. By adding another convolution engine (3 MAC units) in Conv2, this block splits the original array into two arrays (each 34 long because padding is needed for convolution) and performs two convolution processes in parallel as shown in Fig. 12. The convolution results are then combined to obtain the final array. This will reduce the Conv2 delay to 30 clock cycles, which simply means no lag in the processing pipeline shown in Fig. 8. By changing resource allocation in the convolution clocks, the delay is uniformly set to 30 clock cycles. MAC resource allocation for Conv1-3 and Fused Processing Blocks 1-3 are {6, 6, 6} and {44, 33, 30} respectively.

TABLE II
Resources used in FPGA.

| Block Name | ALMs | ALUTs | Dedicated Logic Registers | DSP Blocks |
|---|---|---|---|---|
| Signal Memory | 3931 | 2258 | 14 | 0 |
| Convolution Block 1 | 294 | 173 | 1196 | 3 |
| Convolution Block 2 | 472 | 163 | 1279 | 2 |
| Convolution Block 3 | 275 | 129 | 836 | 2 |
| Fused Processing Block 1 | 861 | 811 | 1680 | 10 |
| Fused Processing Block 2 | 771 | 796 | 943 | 8 |
| Fused Processing Block 3 | 912 | 844 | 482 | 5 |
| Classifier | 58 | 108 | 63 | 2 |
| Scoreboard | 35 | 49 | 28 | 0 |
| Used Resources | 7609 | 5331 | 6521 | 32 |
| Used Resources* | 3643 | 3024 | 6479 | 32 |

\* Without memory and score board.

TABLE III
Power consumption by hierarchy.

| Block Name | Power Consumption |
|---|---|
| Signal Memory | 0.04mW |
| Convolution Block 1 | 0.24mW |
| Convolution Block 2 | 0.16mW |
| Convolution Block 3 | 0.18mW |
| Fused Processing Block 1 | 0.79mW |
| Fused Processing Block 2 | 0.69mW |
| Fused Processing Block 3 | 0.55mW |
| Classifier | 0.02mW |
| Scoreboard | 0.00mW |
| Total Hardware Design | 2.67mW (0.57mW at its own hierarchy) |

TABLE IV
Power consumption by hardware resources.

| Hardware Resources Type | Power Consumption |
|---|---|
| DSP Block | 1.40mW |
| Combinational cell | 0.27mW |
| Clock enable block | 0.32mW |
| Register cell | 0.53mW |
| I/O | 0.15mW |
| Total Hardware Design | 2.67mW |

IV. FPGA AND IMPLEMENTATION RESULTS

A. Resources and Power Consumption

Table II summarizes the utilized resources in the optimized 1-D CNN. As observed, fused processing blocks occupy more resources than convolution blocks. The DSP resources are mainly allocated to fused processing blocks because they perform highly parallel projection tasks. Signal memory block consumes plenty of adaptive logic modules (ALMs) and combinational adaptive look-up tables (ALUTs) for data storage.

The power consumption statistics come from the power analyzer tool in QuartusII, where a clock with 2.5MHz frequency is applied to simulate its power consumption. From the tool, an overall power consumption of 2.67mW is estimated from a 1.1 V supply voltage.

Table III demonstrates the hierarchy of power consumption in this hardware design. Compared to convolution blocks, fused processing blocks consume almost 3X higher power consumption. The power consumption by fused processing block 1, 2 and 3 are 0.79mW, 0.69mW and 0.55mW respectively. Besides, the power consumption of the convolution blocks 1, 2 and 3 are 0.24mW, 0.16mW, 0.18mW respectively. This trend is consistent with the utilized resources, such as registers and DSPs in each block.

Table IV summarizes the the power consumption of the processing pipeline by hardware resources. The DSP block is the most power-consuming, while combinational cells and I/O consume the least amount of power. In the following sections, various testing methodologies are used to evaluate the chip performance under different conditions including confirmation of its successful adaptation providing high clustering accuracy.

B. Dataset Information

To evaluate the performance of the proposed deep spike detector, the Wave_Clus spike bank was used [25]. The database in [25] comprises various average spike waveforms obtained from the neocortex and basal ganglia of humans. To replicate the background noise activity, attenuated spike waveforms selected at random from the data library were incorporated into the generated datasets. There are four datasets in the collected database, each has three spike mean waveforms and provides corresponding spike times and their labels. Besides, the four datasets are categorized according to the different degrees of difficulty (e.g., similarity of spike shape) and the noise levels. The datasets are labeled as C_Easy1_noise, C_Easy2_noise, C_Difficult1_noise, and C_Difficult2_noise, with noise levels represented by standard deviations ($\sigma_N$) of 0.05, 0.1, 0.15, and 0.2. The terms "Easy" and "Difficult" refer to the similarity index between spike shapes in each dataset. Easy1 has also additional noise levels of 0.25, 0.3, 0.35 and 0.4 for further spike sorting performance analysis. Neurons located at a considerable distance from the electrode tips are labeled artefacts, as they reflect spurious neural events that typically show incomplete transitions in polarization or depolarization phases.

C. Classification Performance of the Optimized 1-D CNN

Due to the quantization processing, the data is stored and calculated in the form of fixed-point values. As a result, original samples are converted from FL32 to 10-bit fixed-point values. This format covers the spike variations and ensures that the signals preserve the shapes over time.

The classification performance is evaluated by ModelSim simulation and also deploying the processing pipeline on Cyclone V 5CSEMA5F31C6 FPGA. For example, amongst the 524 batches of testing spike signals, 509 of them are classified correctly by both methods. This results in a hardware classification accuracy of 97.14%.

TABLE V
Comparison of classification accuracy (CAcc) of the proposed deep spike detection on FPGA versus Wave_clus [26] and SpikeDeep-classifier [15] on simulated dataset.

| Dataset | $\sigma_N$ | Spikes | [a]Wave_Clus | [b]SpikeDeep-classifier | [c]Proposed (FPGA) |
|---|---|---|---|---|---|
| Easy 1 | 0.05 | 3514 | 98.82 | 99.07 | 99.43 |
|  | 0.10 | 3522 | 98.81 | 99.32 | 99.77 |
|  | 0.15 | 3477 | 98.72 | 99.10 | 99.43 |
|  | 0.20 | 3474 | 98.55 | 98.85 | 99.44 |
|  | 0.25 | 3298 | 97.32 | 97.13 | 98.11 |
|  | 0.30 | 3475 | 89.50 | 96.75 | 97.18 |
|  | 0.35 | 3534 | 82.12 | 93.34 | 95.83 |
|  | 0.40 | 3386 | 82.60 | 94.72 | 95.37 |
| Easy 2 | 0.05 | 3410 | 96.88 | 96.84 | 98.16 |
|  | 0.10 | 3520 | 91.62 | 91.30 | 94.59 |
|  | 0.15 | 3411 | 91.30 | 92.24 | 93.45 |
|  | 0.20 | 3526 | 84.52 | 83.32 | 90.33 |
| Difficult 1 | 0.05 | 3383 | 89.10 | 96.79 | 97.32 |
|  | 0.10 | 3448 | 93.44 | 93.24 | 95.54 |
|  | 0.15 | 3472 | 66.95 | 70.64 | 87.14 |
|  | 0.20 | 3414 | 75.19 | 67.70 | 85.37 |
| Difficult 2 | 0.05 | 3364 | 93.52 | 95.57 | 98.26 |
|  | 0.10 | 3462 | 94.05 | 83.65 | 96.48 |
|  | 0.15 | 3440 | 83.16 | 81.86 | 93.76 |
|  | 0.20 | 3493 | 56.17 | 77.85 | 86.13 |
| Average CAcc |  |  | 88.15 | 90.46 | 95.05 |

a) Waveclus [26] uses four-level Haar wavelet, Kolmogorov-Smirnov (KS [27]) and superparamagnetic clustering (SPC).
b) SpikeDeep-classifier [15] is based on two CNNs, one for channel selection and the other one for artefact removal.
c) The proposed model has great generalization capability in channel selection. It classified only two out of five hundred and seventy-six channels from recorded data using Utah arrays and micro-wires in [28]-[29].

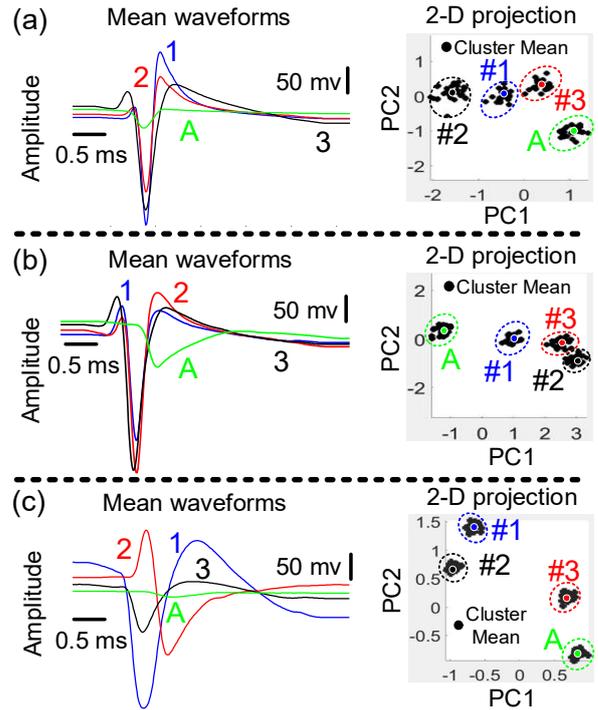

Fig.13. Three different examples of artefact removal are shown in (a)-(c). Color-coded spike mean waveforms corresponding to different neurons. #1 blue, #2 red, #3 black and A in green to represent artefact. 2-D projection of #1, #2, #3 and A. The first two principal components, PC1 and PC2, are used for 2-D projection of spike clusters. The border between the neural events and the artefact fully separated.

This is very close to the derived 99.81% accuracy in MATLAB. One possible explanation for the slight accuracy decrease is the accumulated precision loss. In the MATLAB model, data is still computed in floating-point format, where even a tiny value can be represented. When performing fixed-point number operations in hardware, any value that falls below the current format's precision will be disregarded. While this loss of value may be insignificant when only a few operations are conducted, it can accumulate over time, leading to an increasingly detrimental impact on the overall process as computations progress.

The Classification delay is 42 clock cycles, which is 16.8μs, considering a 2.5MHz clock frequency. The main reason is that each hardware block has to spend several clock cycles on handshaking and state transfer. Also, some blocks have complex control logic for internal resource reuse. These blocks will spend more time on the internal states for register operations, which introduces more cycle lags.

### D. Deep Spike Detection Performance Analysis

FPGA implementation of deep spike detection was tested using Cyclone V 5CSEMA5F31C6 with a MATLAB/Simulink interface to a PC [30]. It embeds two optimized 1-D CNNs into the conventional spike processing pipeline for the selection of the active neural channels and the removal of artefacts from the selected channels, each has an input layer that accepts a 1-D array with a length of 66 samples (duration of 2.5 ms). To extract the most abstract features, the detected spike events are transferred to the PC. Spike waveforms are then transformed to fewer dimensions using principal component analysis (PCA) [31] and fed into the K-means algorithm to identify the classes. Classification accuracy (CAcc) is evaluated by (TPCC/ NTS) ×100%, where TPCC is the number of truly detected and correctly classified spikes and NTS is the number of truly detected spikes. NTS = DTS − (FPS + MS), where DTS is the number of detected spikes, FPS is the number of false alarm spikes due to noise or overlapping spikes, and MS is the number of missed spikes. Table V shows the classification accuracy in comparison with Wave_clus [26] and SpikeDeep-classifier [15]. The proposed deep spike detection method on FPGA demonstrated an accuracy greater than 95%, which shows 6.86% and 4.45% improvement over Wave_clus [26] and SpikeDeep-classifier [15] respectively. Both Wave_clus [26] and SpikeDeep-classifier [15] are developed in software, whereas the proposed deep spike detector is implemented on an FPGA. For visualization purposes, Fig.13 shows the artefact-free classification using label prediction and the PCA-based projection of waveforms onto two-dimensional (2-D) feature space. In all shown cases in Fig.13 (a)-(c), spike events and artefacts are sufficiently separated (i.e., the projection borders are clear).

Simulation suggests a power dissipation of 5.6mW when running at 10-bit resolution and 2.5 MHz operating frequency in the streaming phase.

Downscaling factor (DF) [32] is a mapping process to estimate the dynamic power characteristics of the spike processor in different technologies. The key player in the downscaling process is the supply voltage (Vsup) of different technologies, therefore DF can be expressed as $DF_{Base/Scaling} =$

TABLE VI
The features and outcomes of different spike sorting systems implemented on FPGAs.

| Work | Algorithm | Device | Regs | LUTs | DSP utilization | Freq (MHz) | Clustering Latency | Sorting Accuracy |
|---|---|---|---|---|---|---|---|---|
| Ours | Deep spike detection[a] | Cyclone V | 12958 | 6048 | 64 | 2.5 | 16.8μs [b] | 95.05%[c] |
| [33] | Templated Matching | Virtex-6 | 4880 | 6635 | 5 | 122 | 0.55μs | 90.05% |
| [34] | OSort | Virtex-6 | 8444 | 16472 | 130 | 123 | 0.25μs | 87% |
| [35] | PNN | Virtex-6 | 3936 | 13776 | 54 | 100 | 6.7μs | 93.83% |
| [36] | Hebbian | Spartan-6 | 8904 | 6678 | – | – | 0.96μs | 95% |
| [37] | BOTM | Virtex-6 | 29000 | 190000 | – | – | 2.65ms | – |
| [38] | OSort | Virtex-5 | 16245 | 23567 | 29 | 100 | 11.1ms | – |
| [39] | OSort | Zynq-7000 | 12150 | 14037 | 120 | 101 | 179.4 μs | – |

a) Precision, Throughput (GOP/s) and DSP efficiency (GOP/s/DSPs) are 10bits fixed, 74.4 and 0.42 respectively.
b) 16.8μs pipeline processing delay for real-time and artefact-free spike sorting using handshaking protocol.
c) PCA and K-means are performed off-chip by transferring detected spike events to PC [30].

$(V_{SUP})^2_{Base} / (V_{SUP})^2_{Scaling}$. For evaluation of the spike processor in this work from base technology (1.1 V supply used in FPGA board to scaling technology (65 nm, 0.27 V), both operating at the same frequency, $DF|_{Base/Scaling}$ is calculated at 17. This scaling predicts the deep spike detector power consumption to 329μW in 65nm. This shows compatibility of the proposed design with the implantable brain processing and closed-loop applications. The proposed design is set to be implemented and validated in future work using 65nm technology.

*E. Comparison with the Published Works*

Table VI presents the features and performance outcomes of the proposed deep spike detection spike sorting, in comparison to previously published research conducted on FPGAs. Valencia et. al. [33]-[34] presented template matching and OSort-based clustering modules. Deep spike detector uses more registers and DSP compared to [34], however it provides higher 5% higher sorting accuracy at 2.5 MHz. Note that the deep spike detector offers artefact-free spike sorting, as artefacts will adversely affect the spike sorting accuracy. Calculated sorting accuracy in [33]-[39] assigns equal weights to the artefacts and authentic spike waveforms which results in substantial error in calculating sorting performance. The design outlined in [35] introduces a spike sorting system that employs a probabilistic neural network (PNN). It incorporates a tailored floating-point numerical format comprising 1 sign bit, 8 exponent bits, and 7 fraction bits. The deep spike detector demands comparable resources to [35], while also delivering superior sorting performance. The work in [36] presents a real-time spike sorting system that uses Hebbian learning to implement PCA for projecting input spikes to features of interest. They present results for 16-bit word length, as well as modelling their designs using MATLAB's fixed-point toolbox. Their target device is an Xilinx Spartan-6 FPGA, and their resource utilization was reported based on the reconfigurable slices. It is shown that the deep spike detector implementation uses higher registers and lower LUTs. Note that the design in [36] only reports the implementation results of the Hebbian Eigenfilter hardware. The reported spike sorting accuracy in [36] is 95% and 87% when sorting spikes from three and four neurons, respectively. The design presented in [37] has implemented the Bayes optimal template matching (BOTM) algorithm for spike sorting on a Virtex-6 FPGA. While the maximum operating frequency has not been reported, the sorting latency was stated as 53 sampling cycles at a 20 kHz sampling rate. The BOTM also uses large number of registers and LUTs. The design in [38] is intended for high-speed data processing of neural recordings on a workstation (offline), and the FPGA communicates with the workstation via a PowerPC processor on the interface board. The latency of 11 ms in Table. VI is assumed with 24 kHz sampling rate. This is the worst-case sorting latency, and correlates to 266 clock cycles. Proposed deep spike detector achieves a significant reduction in real-time sorting latency (16.8μs) and 95.05% sorting accuracy.

V. CONCLUSION

This paper proposed a two-layer framework for efficient and real-time hardware implementation of an artefact-free spike sorting. Simulation layer applies structured and non-structured compression techniques including pruning, network projection and quantization to optimize the overall implementation cost of the 1-D CNN. Having applied the simulation layer, a compact model with 99.81% classification accuracy, 41X lower number of parameters and 334X lower memory usage was achieved. Following the simulation layer, hardware techniques including a customized MAC engine, novel fused layers and resource allocation are proposed for a power-efficient system allowing the process of spike waveforms in real-time. Optimized 1-D CNN loses only a small amount of accuracy, while greatly reduces computational complexity and processing latency. It only takes 16.8μs to perform classification on the Cyclone V 5CSEMA5F31C6 FPGA evaluation platform at 2.5MHz. The FPGA prototype provides 97.14% accuracy on a standard dataset and consumes 2.67mW from a 1.1 V supply voltage. An accuracy of 95.05% is achieved with a power of 5.6mW when the deep spike detector is implemented using two optimized 1-D CNNs on the Cyclone V FPGA. Compared to the recently published works, our design can operate at higher frequencies, while minimizing the processing latency, which is ideal for real-time analysis of single-unit activity. Future works will focus on further reduction of computational cost of the deep spike detector, fabrication of the deep spike detector in a specific technology and multi-channel spike sorting using collected in-vivo data.